\documentclass[prd,aps,preprint,superscriptaddress,nofootinbib]{revtex4-1}
\usepackage{epsfig}
\usepackage{amsmath}

\begin{document}

\title{Back-to-Back Correlations of Di-hadrons in dAu Collisions at RHIC}
\author{Anna Stasto}
\affiliation{Department of Physics, Pennsylvania State University, University Park, PA
16802, USA}
\affiliation{RIKEN BNL Research Center, Building 510A, Brookhaven National Laboratory,
Upton, NY 11973, USA}
\affiliation{H. Niewodnicza\'nski Institute of Nuclear Physics, Polish Academy of Sciences, Krak\'ow, Poland}
\author{Bo-Wen Xiao}
\affiliation{Department of Physics, Pennsylvania State University, University Park, PA
16802, USA}
\affiliation{Center for High Energy Physics, Peking University, Beijing 100871, China}
\author{Feng Yuan}
\affiliation{Nuclear Science Division, Lawrence Berkeley National Laboratory, Berkeley,
CA 94720, USA}
\affiliation{RIKEN BNL Research Center, Building 510A, Brookhaven National Laboratory,
Upton, NY 11973, USA}
\affiliation{Center for High Energy Physics, Peking University, Beijing 100871, China}

\begin{abstract}
We perform a complete theoretical analysis of the azimuthal angular correlation 
of two-hadron productions in the forward $dAu$ collisions at RHIC in
the saturation formalism, and obtain a very good agreement with the experimental
data. It is demonstrated that the suppression and broadening of the away side
peak provide a unique signal for the onset of the saturation mechanism at small-$x$
in a large nucleus. We emphasize that future experiments of di-hadron
correlations in $pA$ collisions at both RHIC and LHC, and in $eA$ collisions at
the planned electron-ion collider, shall provide us with a thorough study and understanding of the strong interaction 
dynamics in the saturation regime.
\end{abstract}
\maketitle


{\it Introduction.} Two particle correlations in high energy hadronic scattering
processes have played important roles in studying the strong interaction
QCD dynamics~\cite{Abazov:2004hm,Khachatryan:2011zj}, 
novel phenomena in proton-proton collisions~\cite{Khachatryan:2010gv}, and the 
medium effects in heavy ion collisions~\cite{Adams:2003im,Aad:2010bu,Chatrchyan:2011sx}. 
In particular, the azimuthal correlation
of two particles with large transverse momenta provide critical
evidence of the strong jet quenching effects in the hot dense medium 
created in the heavy ion collisions at RHIC and the LHC. 
An important advantage of these measurements is that they
reveal the physics by the observables themselves, which do not need,
e.g., the proton reference to quantify the phenomena. 
It was suggested in Ref.~\cite{Marquet:2007vb} to study the cold nuclear matter in dense
region by measuring the forward di-jet (di-hadron) production in proton-nucleus collisions, where
the de-correlation of the back-to-back dijet in 
the forward $pA$ collisions can be used to signal the gluon saturation 
at small-$x$ in a large nucleus. This prediction was qualitatively confirmed by the 
STAR~\cite{Braidot:2010ig} and PHENIX~\cite{Adare:2011sc} collaborations from 
di-hadron correlation measurements in $dAu$ collisions 
at RHIC, and have been considered as the best evidence for saturation physics. 
Early attempts~\cite{Albacete:2010pg,Tuchin:2009nf} have been made to understand 
the experimental data quantitatively, where, however, inappropriate approximations
were taken in the calculations. 
In particular, in this paper we  correctly distinguish
the Weizs\"{a}cker-Williams (WW) gluon distribution
from the dipole gluon distribution.
In this work, we will provide, for the first time, a quantitative and thorough
description of the experimental data in the saturation formalism, 
including the large broadening of the angular distribution and 
suppression of the peak for the away-side hadron. 
Our results emphasize the relevance
of the gluon saturation in the kinematic regions covered by the STAR 
and PHENIX collaborations and signify the usage of the two-hadron
correlation as an important tool to investigate the QCD dynamics
in the small-$x$ limit. 

There have been theoretical arguments~\cite{Gribov:1984tu,Mueller:1985wy,McLerran:1993ni,Iancu:2003xm}
which suggest that the gluon distribution saturates at small
Bjorken-$x$. The color-glass-condensate model (CGC), has been proposed to
describe the gluon saturation phenomenon at small-$x$. 
An important feature of this approach is the appearance of the dynamically generated saturation scale $Q_s(x)$ which  separates the dilute and dense partonic regimes.
The experimental data from
HERA are well described with the saturation model calculations,
with the saturation scale  of the order of $\sim 1 \; {\rm GeV}$.
Furthermore, the forward hadron production in
nucleon-nucleus ($pA$) collisions has been systematically studied~\cite{Dumitru:2002qt} in the CGC
formalism, where the unintegrated gluon distributions
(UGDs) are important ingredients to describe the phenomena. 
They unveil the importance of the multiple interaction
effects in the factorization of the hard processes in the
small-$x$ calculations. Nevertheless, there exists much more 
interesting dynamics~\cite{Dominguez:2010xd} in saturation physics which 
can only be explored by di-jet or di-hadron production processes
as we will demonstrate in the following calculations. 

In this paper, we focus on two-particle production in the forward direction of $pA$ ($dAu$ at RHIC)
collisions, 
\begin{equation}
p+A\to h_1+h_2+X \ , \label{pa}
\end{equation}
where two hadrons $h_1$ and $h_2$ with large
momenta are produced. The above process is sensitive to the gluon
distributions at small-$x$ in the nuclear target.
In order to correctly take into account the multiple 
interaction effects, we follow the CGC
framework to calculate the two particle production~\cite{Dominguez:2010xd}. 
An effective $k_t$ factorization
can be established for this process in the back-to-back
correlation limit, and the
differential cross sections can be expressed in terms of various
UGDs, which can be related to two fundamental UGDs: the dipole
gluon distribution $xG^{(2)}(x,q_{\perp })$, and the
WW gluon distribution
$xG^{(1)}(x,q_{\perp })$. Only with this effective $k_t$ factorization, 
can one describe all the features (including both broadening and suppression) 
of the STAR~\cite{Braidot:2010ig} and PHENIX~\cite{Adare:2011sc} data systematically.
These results also agree with previous calculations
for two-particle production in $pA$
collisions in the general kinematics region~\cite{Blaizot:2004wv, JalilianMarian:2004da}. 

In the RHIC experiments, the di-hadron correlations are measured by the
coincidence probability $C(\Delta \phi) ={N_{\text{pair}}(\Delta \phi)}/{N_{\text{trig}}}$, where
$N_{\text{pair} }(\Delta \phi)$ is the yield of two forward
$\pi^0$ which includes a trigger particle with a transverse
momentum $p_{1\perp}^{\text{trig}}$ and an
associate particle with  $p_{2\perp}^{\text{asso}}$ and
the azimuthal angle between them $\Delta \phi$.
We calculate the single and two-particle cross sections and obtain,
\begin{eqnarray}
C(\Delta \phi) =\frac{\int_{|p_{1\perp}|,|p_{2\perp}|}\frac{d\sigma^{pA\to h_1h_2}}{dy_1 dy_2 d^2 p_{1\perp}d^2p_{2\perp}}}
{\int_{|p_{1\perp}|}\frac{d\sigma^{pA\to h_1}}{dy_1 d^2 p_{1\perp}}} \ , \label{cp}
\end{eqnarray}
where the dependence on the rapidities of the two particles is implicit.

{\it Single inclusive cross section.}
Let us first discuss the single inclusive hadron production.
The leading-order single inclusive cross section~\cite{Dumitru:2002qt} in $pA$
collisions is given by the product of the integrated parton 
distributions of the projectile proton and the unintegrated gluon
distributions of the target nucleus:
\begin{eqnarray}
\frac{d\sigma^{pA\rightarrow hX}}{d^2 b \,d^{2}p_{\perp }\,dy_h}&=&
\int_{z_h}^1 \frac{dz_1}{z_1^2}  \left[D_{h/q}(z_1)
x_p q_f(x_p)F_{x_g}(k_{\perp }) \nonumber\right.\\
&&\left.+\;x_p g_f(x_p) \tilde{F}_{x_g}(k_{\perp})  
D_{h/g}(z_1)\right] \ , \label{single}
\end{eqnarray}
where the sum over quark flavor is implicit, $b$ represents the impact parameter in $pA$ collisions,
$p_\perp$ and $y_h$ are transverse momentum and rapidity of the hadron,
$q(x_p)$ and $g(x_p)$ are integrated quark and gluon distributions
from the projectiles, $D(z)$ the associated fragmentation functions with
$p_{\perp}=z_1k_{\perp}$, $x_p={p_{\perp}}e^{y_h}/{z_1\sqrt{s}}$
and $x_g={p_{\perp}}e^{-y_h}/{z_1\sqrt{s}}$. The dipole gluon
distributions $F_{x_g}(k_\perp)$ and $\tilde F_{x_g}(k_\perp)$
are Fourier transform of the dipole scattering amplitude
in the fundamental and adjoint representations, respectively.
In particular, $F_{x_g}(k_\perp) \propto x_gG^{(2)}(x_g,k_\perp)/k_\perp^2$.
In terms of the numerical study, we are able to describe the forward
single hadron production cross sections measured by
both BRAHMS and STAR up to $p_{\perp}=3.0 \, \text{GeV}$ with a K-factor about 0.8
for $y_h=2.0$ and 0.5 for $y_h=3.2$.
In this numerical evaluation, we follow the NLO sets of MSTW parametrizations \cite{Martin:2009iq}
for the parton distributions and DSS parametrizations \cite{deFlorian:2007aj}
for the fragmentation functions~\footnote{A recent next-to-leading order
calculation for inclusive hadron production suggests that the 
appropriate choice for the factorization scale to be around the saturation scale~\cite{Chirilli:2011km}. 
We have followed this choice in our calculations.}.

{\it Two-particle production in forward $pA$ collisions.}
Two-particle production contains the correlated and uncorrelated contributions,
\begin{eqnarray}
d\sigma^{(pA\to h_1 h_2)}=d\sigma_{\rm corr.}^{(pA\to h_1h_2)}
+d\sigma_{\rm uncorr.}^{(pA\to h_1h_2)} \ .
\end{eqnarray}
The correlated hadron production comes from the partonic $2\to 2$ processes,
where these two particles are back-to-back correlated and form the away
side peak in the azimuthal angular distribution ($\Delta \phi=\pi$). The near side
correlation comes from the particle decay or the same jet fragmentation
if they are at the same rapidity. In this letter, we will focus on the back-to-back
correlation region, namely the away side peaks. According to Ref.~\cite{Dominguez:2010xd}, we can write down
the differential cross section for the two-particle production
in the back-to-back correlation limit,
\begin{eqnarray}
&&\frac{d\sigma_{\rm corr.}^{(pA\rightarrow h_1h_2)}}{dy_{h_1}dy_{h_2}d^{2}p_{1%
\perp }d^{2}p_{2\perp }} =\int\frac{dz_1}{z_1^2}\frac{dz_2}{z_2^2}
\frac{\alpha _{s}^{2}}{\hat{s}^{2}}\left[
x_{p}q(x_{p}) \mathcal{F}_{qg}^{(i)}\nonumber\right.\\
&&~\times H_{qg}^{(i)}
\left(D_{h_1/q}(z_1)D_{h_2/g}(z_2)
+D_{h_2/q}(z_1)D_{h_1/g}(z_2)\right)\nonumber\\
&&~\left.+x_{p}g(x_{p})\mathcal{F}_{gg}^{(i)}H_{gg}^{(i)}D_{h_1/g}(z_1)D_{h_2/g}(z_2)\right] \ ,  \label{dijet}
\end{eqnarray}%
where $x_g=x_1e^{-y_1}+x_2e^{-y_2}$ and $x_p=x_1e^{y_1}+x_2e^{y_2}$
with $x_i={|k_{i\perp }|}/{\sqrt{\hat s}}$ and $k_{i\perp}=p_{i\perp}/z_i$,
$\mathcal{F}^{(i)}$ and $H_{}^{(i)}$ are various UGDs
and the associated hard coefficients, respectively. Their expressions
can be found in Ref.~\cite{Dominguez:2010xd}.
The partonic center of mass energy squared $\hat s$ is defined
as $\hat{s}={P_{\perp }^{2}}/{z(1-z)}$ with
$P_{\perp }=\left( k_{1\perp }-k_{2\perp }\right)/2 $ and $z=x_1e^{y_1}/x_2 e^{y_2}$.
In the CGC calculations~\cite{Dominguez:2010xd}, 
$\tilde P_\perp=\left( 1-z\right) k_{1\perp }-zk_{2\perp }$ also enters in the
hard coefficients, which equals to $P_\perp$ in the correlation limit. The difference between
$P_\perp$ and $\tilde P_\perp$ will be used to estimate the theoretical uncertainties in 
the following calculations.
In the typical kinematics of the forward collisions at RHIC, we find that 
$x_p\sim0.1$ and $x_g \leq 10^{-3}$, where
both the quark initiated processes ($q\to qg$ channel) and gluon initiated
processes ($g\to gg$) contribute.

Comparing the above equation to Eq.~(3) of Ref.~\cite{Albacete:2010pg}, one immediately
finds notable differences between the results.
In particular in Ref.~\cite{Albacete:2010pg} the only channel calculated was $q\to qg$.
Moreover, in this channel our results do not agree with results in Ref.~\cite{Albacete:2010pg}
since in the latter work the contributions from the WW gluon distribution  were not taken into account.  These contributions are essential in order to 
 reproduce correctly  the collinear factorization results for dijet production in the dilute
limit. 

The unintegrated gluon distributions in Eq.~(\ref{dijet}) 
are largely un-explored, in particular, for those related to the WW
gluon distribution. The energy evolution is important to understand 
their behavior depending on $x_g$, of which for the dipole gluon distribution, the 
Balitsky-Kovchegov (BK) evolution, has been well studied~\cite{Iancu:2003xm} and
demonstrated the so-called geometric scaling~\cite{Stasto:2000er} in the solution.
The scaling was found to be related to the traveling wave solutions~\cite{Munier:2003vc,
Mueller:2002zm} of the BK evolution. 
The energy evolution equation for the WW gluon distribution has recently been
systematically investigated~\cite{Dominguez:2011gc}.
An important result from these studies is the geometric scaling similar 
to the dipole gluon distribution. 
Therefore, as a first step, we can parametrize these gluon distributions 
from a model calculation, and include the energy dependence by assuming the
geometric scaling and $x_g$-dependence of the saturation scale.  
In the following, we adopt the Golec-Biernat Wusthoff model~\cite{GolecBiernat:1998js}
for the dipole gluon distribution
which successfully describes the low-$x$ DIS structure functions at HERA,  
then extend it to the WW gluon distribution and 
include the nuclear dependence by modifying the saturation scale as~\cite{Iancu:2003xm} 
\begin{equation}
Q_{sA}^2=c(b)A^{1/3}Q_s^2(x)\ , \label{cb}
\end{equation}
where $c(b)$ represents the profile function of nucleus depending on the 
impact parameter $b$ of the collision,  $Q_s^2(x)=
Q_{s0}^2 (x/x_0)^{-\lambda}$ with $Q_{s0}=1\,\text{GeV}$, $x_0=3.04\times 10^{-3}
$ and $\lambda=0.288$ follow GBW parameterizations~\cite{GolecBiernat:1998js}.
The profile function $c(b)$ is closely related to the centrality of the $pA$ (or $dA$) collisions. 
Central collisions give large value of $c(b)$, while peripheral collisions correspond to small profile function.

We would like to emphasize that the GBW model is not sufficient to describe 
the UGDs in the region that $k_\perp$ is much larger than $Q_s$. However,
for the forward $pA$ collisions, the saturation scale $Q_s$
is large enough to cover most of the kinematics where $k_\perp$ is 
around $Q_s$ and we will be able to well describe the experimental data.
For $pp$ collisions, we have to either modify the GBW model or include the broadening
effect from the fragmentation function to describe the experimental data.

{\it Double parton scattering contribution.}
Now, we turn to the un-correlated two-particle production in the process (\ref{pa}). 
This part mainly comes from two independent hard scatterings, which is referred as  
double parton scattering (DPS) contributions (see
recent developments~\cite{Gaunt:2010pi,Diehl:2011tt,Strikman:2010bg,Gaunt:2009re,Blok:2010ge}).
It has been pointed out in Ref.~\cite{Strikman:2010bg}
that the DPS may exceed the single parton scattering contribution
in the forward $pA$ collisions. 
Following these ideas, we estimate its contributions in $pA$ collisions in the saturation formalism.
In particular, the multiple interactions from the nuclei side has been
taken into account in the CGC factorization formalism~\cite{Iancu:2003xm}. 
For the proton side, we follow a simple parametrization for the double
parton distribution: ${\cal D}_{p}^{ij}(x_p,x_p')={\cal C}(x_p,x_p')f_{i}(x_p)\times f_{j}(x_p')$
with ${\cal C}\approx 1$, where $i$ and $j$ represent the two partons from the nucleon
which participate the hard scattering, $x_p$ and $x_p'$ for their momentum fractions.
The final expression reads as
\begin{eqnarray}
&&\frac{d\sigma_{\rm uncorr.}^{(pA\rightarrow h_1h_2)}}{d^2bdy_{h_1}dy_{h_2}d^{2}p_{1%
\perp }d^{2}p_{2\perp }} =\int\frac{dz_1}{z_1^2}\frac{dz_2}{z_2^2}D(z_1)D(z_2)\nonumber\\
&&~~~~~\times 
\sum_{ij}x_{p}f_i(x_{p})x_{p}'f_j(x_{p}')F^{(i)}_{x_g}(k_{1\perp})F^{(j)}_{x_g'} (k_{2\perp}) \ ,\label{dps}
\end{eqnarray}
where $ij$ represent flavors of the partons from the nucleon and the associated UGDs from
nuclei, and $x_p$ and $x_g$ are determined by the kinematics of the two hard
scatterings. 

An important feature of the above DPS contribution is that the two hard scatterings 
are independent to each other at the leading order approximation~\cite{Gaunt:2010pi,Diehl:2011tt,Strikman:2010bg}. 
Therefore, the two particles in the final state are un-correlated, and their azimuthal angle 
distribution will be flat. This leads to the so-called pedestal contribution 
in the experimental measurements.

\begin{figure}[tbp]
\begin{center}
\includegraphics[width=7cm,height=5cm]{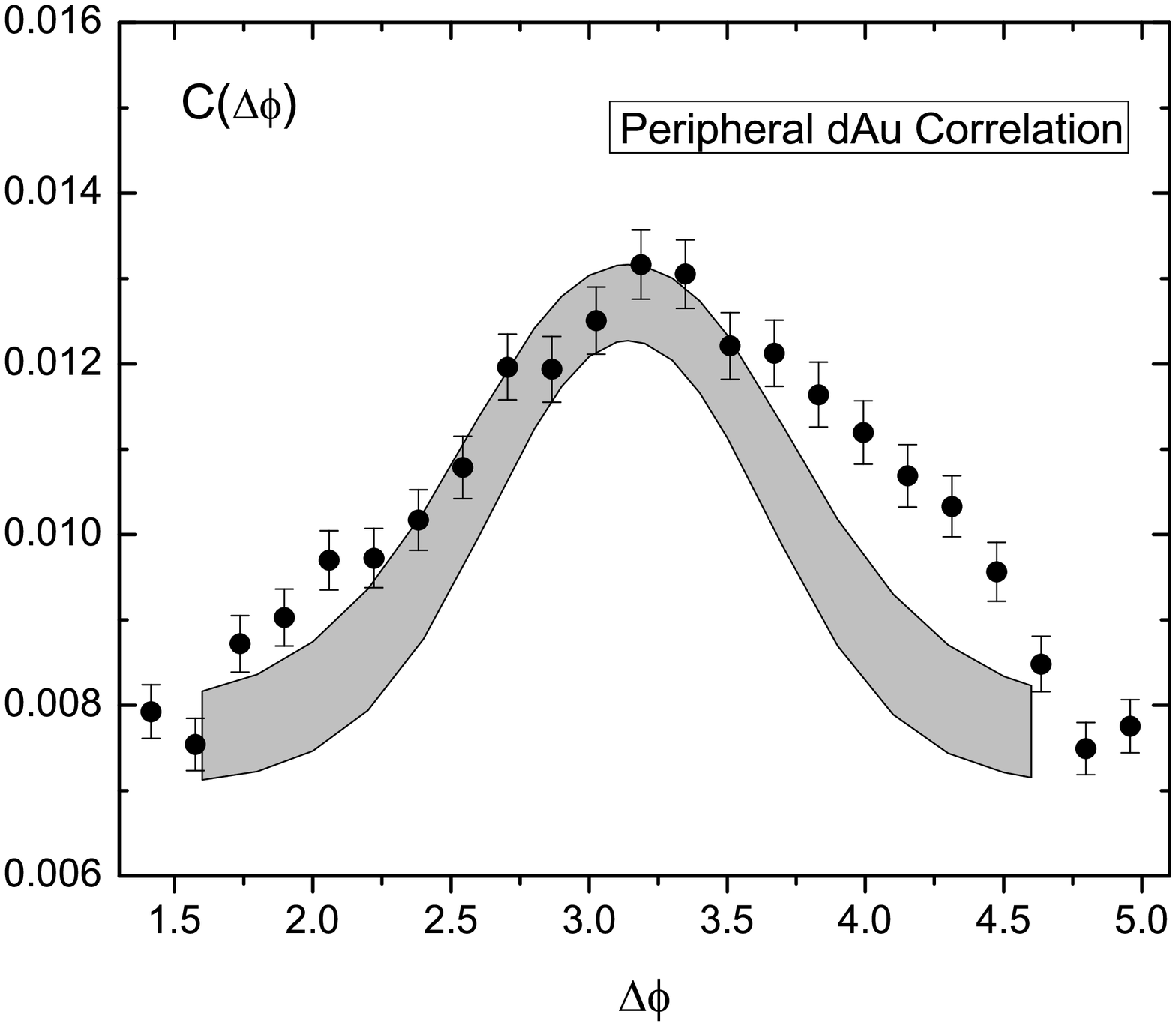}
\includegraphics[width=7cm,height=5cm]{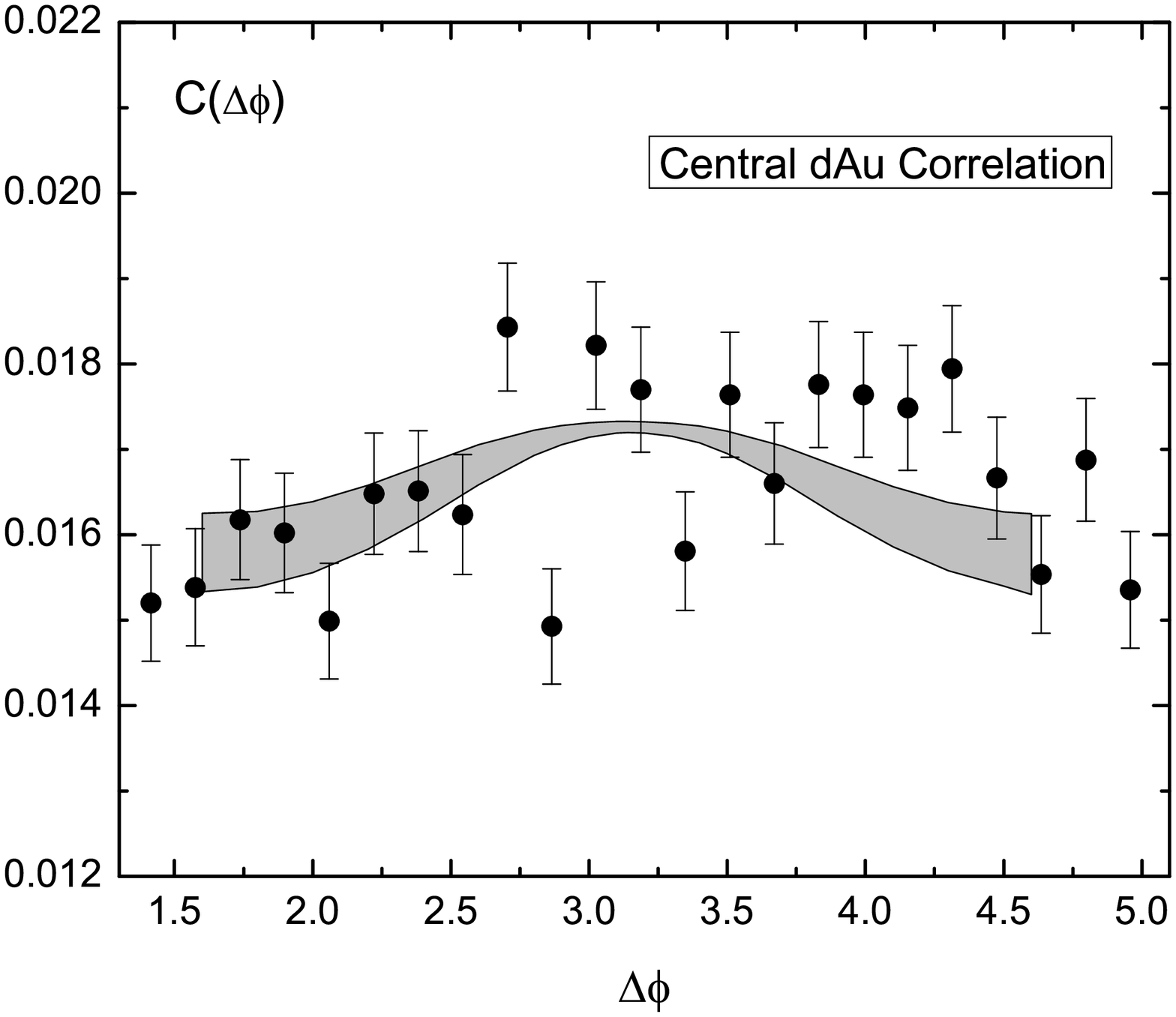}
\end{center}
\caption[*]{The forward di-pion correlations $C(\Delta \phi)$ of Eq.~(\ref{cp}) at $y_1\sim y_2\sim 3.2$ 
in peripheral and central $dAu$ collisions compared to the preliminary data 
from the STAR collaboration~\cite{Braidot:2010ig}.
Centrality definition follows Ref.~\cite{Braidot:2010ig}, where the average 
impact parameters are found around 6.7fm and 2.7fm accordingly.
The grey error band comes from using $P_\perp$ or $\tilde P_\perp$ in
the hard coefficients in Eq.~(\ref{dijet}).}
\label{rhic}
\end{figure}

{\it Comparison with the experimental data.} With the above
formulas, we are ready to compare to the experimental data
on the two-particle correlation measurements
in the forward $dAu$ collisions at RHIC. 
Before we do that, we would like to emphasize that the saturation
scale plays a key role in describing the correlation $C(\Delta\phi)$ 
of the away side peak, including both broadening and the suppression.
First, the width of the away side peak will increase with the saturation scale
because of the broadening effects. Quantitatively,
the effective $k_t$-factorization formula of Eq.~(\ref{dijet})
lead to stronger broadening effects compared to the 
naive $k_t$-factorization calculations. This is because the various 
gluon distributions contain the convolution of the UGDs and 
will enhance the broadening. Without this enhancement, 
we can not describe the broadening effects. In particular, 
when the saturation scale reaches the transverse momenta
of the dijet, the away side peak will almost disappear as 
indicated in the experimental data for the central collisions
at RHIC and the theory calculations as well.

Second, the magnitude of the correlation $C(\Delta \phi)$ is also
sensitive to the saturation scale $Q_s$. In particular, larger $Q_s$
push the dipole gluon distribution to larger transverse momentum, which leads to
single particle production (\ref{single}) increasing with $Q_s$. The correlated two-particle
production cross section (\ref{dijet}), however, decreases with $Q_s$ for the 
same reason. Therefore, the correlated contribution to $C(\Delta \phi)$ decreases
accordingly. Our numeric evaluation also supports this conclusion.
On the other hand, the un-correlated two particle production cross
section (\ref{dps})  roughly depends on the product of two single particle
cross sections. Therefore, its contribution increases more rapidly with $Q_s$
than that of single particle cross section. The consequence is that the
pedestal contribution increases with $Q_s$.

All these features are evident when we compare to the STAR data~\cite{Braidot:2010ig}. 
As an example, we show in Fig.~1 the results for $p_{1\perp}^{\rm trig}>2\,{\rm GeV}$
and $1\,{\rm GeV}<p_{2\perp}^{\rm asso}<p_{1\perp}^{\rm trig}$ at $y_1\sim y_2\sim 3.2$ in 
the peripheral and central collisions, respectively.
In our calculations, we assume a fixed strong coupling constant 
$\alpha_s =0.35$. The saturation scale $Q_s$
($c(b)$ in Eq.~(\ref{cb})) is the only parameter to fit the data, for
which we found $c(b)=0.45,~0.56,~0.85$ for the peripheral, minimum bias,
and central collisions, respectively. These parameters are consistent
with $c(0)\approx 0.9$ and the centrality dependence of the nuclei 
profile for these collisions, by using either the hard sphere model
$c(b)=c(0)\sqrt{1-b^2/R_A^2}$ or the Wood-Saxon model. The pedestal
contributions to $C(\Delta\phi)$ are found to be around 0.016 and 0.018
for the peripheral and central collisions, which are in rough agreement
with the experimental measurements. In the plots of Fig.~1, in order to 
better compare the results, we used the experimental extractions of pedestal contributions.
The grey error band of the theory calculations comes from the difference
between $P_\perp$ and $\tilde P_\perp$ used in the hard coefficients in Eq.~(\ref{dijet}).
Similar results are also found when we compare to the correlation measurements 
from PHENIX collaboration.

It is important to note that in the central $dAu$ collisions the disappearing 
of the away-side peak in the di-hadron production indicates that the saturation
scale is the same order as the hard probe of the jet transverse momentum, which is a clear signal of the onset of the saturation mechanism for this observable.
From the kinematics of this collision, we conclude that the saturation scale
$Q_s$ reaches at $\sim 2 \, {\rm GeV}$ at $x_g\sim 6\times 10^{-4}$ in the center of the gold
nucleus with jet transverse momentum $k_\perp\sim 3 \, {\rm GeV}$ at rapidity $3.2$.
We hope that the future measurements at RHIC and LHC will provide more information
and help to map out the complete phase structure of the cold nuclei at small-$x$.
We want also to emphasize that the above conclusion is very general and independent of the model
we used for the UGDs. As mentioned above, the GBW model
captures the main features for the UGDs at the transverse momentum
around the saturation scale in which most of the data exist.

\begin{figure}[tbp]
\begin{center}
\includegraphics[width=8cm,height=6.0cm]{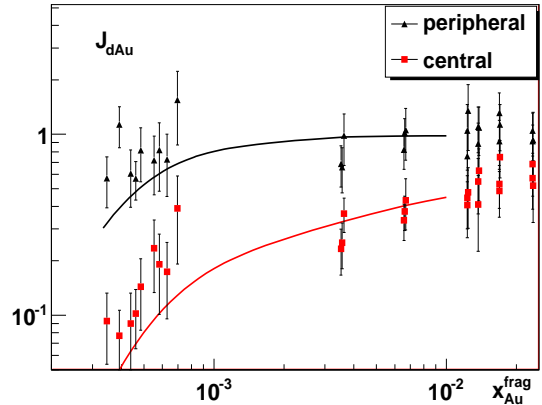}
\end{center}
\caption[*]{The nuclear suppression factor $J_{dA}$ of Eq.~(\ref{jda}) for the two particle
production in $dAu$ collisions as function of $x_{\rm Au}^{\rm frag}$. The experimental
data from PHENIX~\cite{Adare:2011sc} and theory calculations are pedestal contributions subtracted.}
\label{phenix}
\end{figure}

Meanwhile, the PHENIX collaboration has also reported the nuclear suppression 
factor $J_{dA}$ defined as
\begin{equation}
J_{dA}=\frac{1}{\langle N_{\rm coll}\rangle}\frac{\sigma_{dA}^{\rm pair}/\sigma_{dA}}
{\sigma_{pp}^{\rm pair}/\sigma_{pp}} \ ,\label{jda}
\end{equation} 
where $\sigma^{\rm pair}$ is the cross section of two-particle production in $dA$ and $pp$ collisions
with the pedestal contributions subtracted. $\sigma_{dA}$ and $\sigma_{pp}$ are the cross sections for the full event selection. In the absence of gluon saturation and nuclear effects, the dihadron cross-sections are expected to scale with $\langle N_{\rm coll}\rangle$. Therefore, $J_{dA}$ should be equal to unity in the dilute regime and suppressed in the dense regime.
The experimental kinematic variable 
$x_{\rm Au}^{\rm frag}=(p_{1\perp}e^{-y_1}+p_{2\perp}e^{-y_2})/\sqrt{s}$ 
has been used to represent the $x$-dependence~\cite{Adare:2011sc}. 
In Fig.~2, we calculated $J_{dA}$ as function of $x_{\rm Au}^{\rm frag}$
for a typical transverse momentum $p_{1\perp}=p_{2\perp}=1.0\,{\rm GeV}$ 
with the same rapidity for the two particles, for central collision ($c(b)=0.85$)
and peripheral collision ($c(b)=0.45$), where we have used $\tilde P_\perp$
in the hard coefficients in Eq.~(\ref{dijet}). Similar results are obtained with
choice of $ P_\perp$. 
As a comparison, we list the PHENIX data in Fig.~2, where the data are
for different values of $p_{i\perp}$ and rapidities. From this figure, we 
clearly see that the the suppression of $J_{dA}$ at low $x_{\rm Au}^{\rm frag}$ due to the saturation of the cross sections in $pp$ and $dA$ collisions, 
as also indicated by the experimental data. However, we emphasize that 
$J_{dA}$ depends on the $pp$ reference, for which our model calculations should be
taken with cautions.


In summary, we have carried out a complete numerical study on the
forward dihadron correlations in $dAu$ collisions and found good
agreement with the experimental data from RHIC.
These results demonstrated that the saturation formalism developed 
recently can be used to describe the broadening and suppression 
of the away side peak in the di-hadron production in $pA$ collisions.
This emphasizes that the di-hadron (dijet) correlation provides
a unique signal for the onset of saturation mechanism at small-$x$
in a large nucleus. Future experiments at RHIC, EIC and LHC 
will provide excellent opportunities to thoroughly investigate the QCD dynamics in the 
saturation regime. 

We thank E. Avsar, L. Bland, M. Chiu, F. Dominguez, C.
Marquet, L. McLerran, A. H. Mueller, J.W. Qiu, M. Strikman, R. Venugopalan, W.
Vogelsang and N. Xu for helpful conversations. In particular,
we are grateful to R. Venugopalan for stimulating discussions
and comments. This work was
supported in part by the U.S. Department of Energy under the
contracts DE-AC02-05CH11231 and DOE OJI grant No. DE - SC0002145.
We are grateful to RIKEN, Brookhaven National Laboratory and the
U.S. Department of Energy (contract number DE-AC02-98CH10886) for
providing the facilities essential for the completion of this
work.

\end{document}